\documentclass[twocolumn]{aastex631}

\usepackage{xcolor}

\usepackage{threeparttable}
\usepackage{pifont}

\def\lcii{$L_{\rm [CII]}$}

\def\zphot{\ifmmode z_{\rm phot}\else$z_{\rm phot}$\fi}

\def\Oi{[O\,{\sc i}]}
\def\Oiii{[O\,{\sc iii}]}
\def\Cii{[C\,{\sc ii}]}
\def\ltsima{$\buildrel<\over\sim$}
\def\la{\lower.5ex\hbox{\ltsima}~}
\def\gtsima{$\buildrel>\over\sim$}
\def\ga{\lower.5ex\hbox{\gtsima}~}
\def\deg~{$^{\circ}$}

\def\aap{A\&A}

\shorttitle{\Cii\,$158\,\rm{\mu m}$ scan of a star-forming galaxy at $z=6.2078$}
\shortauthors{Fudamoto et al.}

\begin{document}

\title{The Extended \Cii\ under Construction? Observation of the brightest high-z lensed star-forming galaxy at z = 6.2}

\email{yoshinobu.fudamoto@gmail.com}

\author[0000-0001-7440-8832]{Yoshinobu Fudamoto} 
\affiliation{Waseda Research Institute for Science and Engineering, Faculty of Science and Engineering, Waseda University, 3-4-1 Okubo, Shinjuku, Tokyo 169-8555, Japan}
\affiliation{National Astronomical Observatory of Japan, 2-21-1, Osawa, Mitaka, Tokyo, Japan}

\author[0000-0002-7779-8677]{Akio K. Inoue} 
\affiliation{Waseda Research Institute for Science and Engineering, Faculty of Science and Engineering, Waseda University, 3-4-1 Okubo, Shinjuku, Tokyo 169-8555, Japan}
\affiliation{Department of Physics, School of Advanced Science and Engineering, Faculty of Science and Engineering, Waseda University, 3-4-1, Okubo, Shinjuku, Tokyo 169-8555, Japan}

\author[0000-0001-7410-7669]{Dan Coe} 
\affiliation{Space Telescope Science Institute (STScI), 3700 San Martin Drive, Baltimore, MD 21218, USA}
\affiliation{Association of Universities for Research in Astronomy (AURA) for the European Space Agency (ESA), STScI, Baltimore, MD, USA}
\affiliation{Center for Astrophysical Sciences, Department of Physics and Astronomy, The Johns Hopkins University, 3400 N Charles St. Baltimore, MD 21218, USA}

\author[0000-0003-1815-0114]{Brian Welch}
\affiliation{Department of Astronomy, University of Maryland, College Park, MD 20742, USA}
\affiliation{Observational Cosmology Lab, NASA Goddard Space Flight Center, Greenbelt, MD 20771, USA}
\affiliation{Center for Research and Exploration in Space Science and Technology, NASA/GSFC, Greenbelt, MD 20771}

\author[0000-0003-3108-9039]{Ana~Acebron}
\affiliation{Dipartimento di Fisica, Universit{\`a} degli Studi di Milano, Via Celoria 16, I-20133 Milano, Italy}
\affiliation{INAF - IASF Milano, via A. Corti 12, I-20133 Milano, Italy}

\author[0000-0003-4223-7324]{Massimo Ricotti}
\affiliation{Department of Astronomy, University of Maryland, College Park, 20742, USA}

\author[0000-0001-8057-5880]{Nir Mandelker}
\affiliation{Centre for Astrophysics and Planetary Science, Racah Institute of Physics, The Hebrew University, Jerusalem, 91904, Israel}

\author[0000-0001-8156-6281]{Rogier A. Windhorst}
\affiliation{School of Earth and Space Exploration, Arizona State University,
Tempe, AZ 85287-1404, USA}

\author[0000-0002-9217-7051]{Xinfeng Xu}
\affiliation{Center for Astrophysical Sciences, Department of Physics \& Astronomy, Johns Hopkins University, Baltimore, MD 21218, USA}

\author[0000-0001-6958-7856]{Yuma Sugahara}
\affiliation{Waseda Research Institute for Science and Engineering, Faculty of Science and Engineering, Waseda University, 3-4-1 Okubo, Shinjuku, Tokyo 169-8555, Japan}
\affiliation{National Astronomical Observatory of Japan, 2-21-1, Osawa, Mitaka, Tokyo, Japan}

\author[0000-0002-8686-8737]{Franz E. Bauer}
\affiliation{Instituto de Astrof{\'{\i}}sica, Facultad de F{\'{i}}sica, Pontificia Universidad Cat{\'{o}}lica de Chile, Campus San Joaqu\'{i}n, Av. Vicu\~{n}a Mackenna 4860, Macul Santiago, Chile, 7820436} 
\affiliation{Centro de Astroingenier{\'{\i}}a, Facultad de F{\'{i}}sica, Pontificia Universidad Cat{\'{o}}lica de Chile,  Campus San Joaqu\'{i}n, Av. Vicu\~{n}a Mackenna 4860, Macul Santiago, Chile, 7820436} 
\affiliation{Millennium Institute of Astrophysics, Nuncio Monse{\~{n}}or S{\'{o}}tero Sanz 100, Of 104, Providencia, Santiago, Chile} 

\author[0000-0001-5984-0395]{Maru{\v s}a Brada{\v c}}
\affiliation{University of Ljubljana, Department of Mathematics and Physics, Jadranska ulica 19, SI-1000 Ljubljana, Slovenia}
\affiliation{Department of Physics and Astronomy, University of California Davis, 1 Shields Avenue, Davis, CA 95616, USA}

\author[0000-0002-7908-9284]{Larry D. Bradley}
\affiliation{Space Telescope Science Institute (STScI), 3700 San Martin Drive, Baltimore, MD 21218, USA}

\author[0000-0001-9065-3926]{Jose M. Diego}
\affiliation{Instituto de F\'isica de Cantabria (CSIC-UC). Avda. Los Castros s/n. 39005 Santander, Spain}

\author[0000-0001-5097-6755]{Michael Florian}
\affiliation{Department of Astronomy, Steward Observatory, University of Arizona, 933 North Cherry Avenue, Tucson, AZ 85721, USA}

\author[0000-0003-1625-8009]{Brenda Frye}
\affiliation{Department of Astronomy, Steward Observatory, University of Arizona, 933 North Cherry Avenue, Tucson, AZ 85721, USA}

\author[0000-0001-7201-5066]{Seiji Fujimoto}
\affiliation{
Department of Astronomy, The University of Texas at Austin, Austin, TX 78712, USA
}

\author[0000-0002-0898-4038]{Takuya Hashimoto}
\affiliation{Division of Physics, Faculty of Pure and Applied Sciences, University of Tsukuba,Tsukuba, Ibaraki 305-8571, Japan}
\affiliation{Tomonaga Center for the History of the Universe (TCHoU), Faculty of Pure and Applied Sciences, University of Tsukuba, Tsukuba, Ibaraki 305-8571, Japan}

\author[0000-0002-6586-4446]{Alaina Henry}
\affiliation{Space Telescope Science Institute (STScI), 3700 San Martin Drive, Baltimore, MD 21218, USA}
\affiliation{Center for Astrophysical Sciences, Department of Physics and Astronomy, The Johns Hopkins University, 3400 N Charles St. Baltimore, MD 21218, USA}

\author[0000-0003-3266-2001]{Guillaume Mahler}
\affiliation{Institute for Computational Cosmology, Durham University, South Road, Durham DH1 3LE, UK}
\affiliation{Centre for Extragalactic Astronomy, Durham University, South Road, Durham DH1 3LE, UK}

\author[0000-0001-5851-6649]{Pascal A. Oesch}
\affiliation{Departement d'Astronomie, Universit\'e de Gen\'eve, 51 Ch. Pegasi, CH-1290 Versoix, Switzerland}
\affiliation{Cosmic Dawn Center (DAWN), Niels Bohr Institute, University of Copenhagen, Jagtvej 128, K\o benhavn N, DK-2200, Denmark}

\author[0000-0002-5269-6527]{Swara Ravindranath}
\affiliation{Space Telescope Science Institute (STScI), 3700 San Martin Drive, Baltimore, MD 21218, USA}

\author[0000-0002-7627-6551]{Jane Rigby}
\affiliation{Observational Cosmology Lab, NASA Goddard Space Flight Center, Greenbelt, MD 20771, USA}

\author[0000-0002-7559-0864]{Keren Sharon}
\affiliation{Department of Astronomy, University of Michigan, 1085 S. University Ave, Ann Arbor, MI 48109, USA}

\author[0000-0002-6338-7295]{Victoria Strait}
\affiliation{Cosmic Dawn Center (DAWN), Copenhagen, Denmark}
\affiliation{Niels Bohr Institute, University of Copenhagen, Jagtvej 128, Copenhagen, Denmark}

\author[0000-0003-4807-8117]{Yoichi Tamura}
\affiliation{Department of Physics, Graduate School of Science, Nagoya University, Aichi 464-8602, Japan}

\author[0000-0001-9391-305X]{Michele Trenti}
\affiliation{School of Physics, University of Melbourne, Parkville 3010, VIC, Australia}
\affiliation{ARC Centre of Excellence for All Sky Astrophysics in 3 Dimensions (ASTRO 3D), Australia}

\author[0000-0002-5057-135X]{Eros Vanzella}
\affiliation{INAF -- OAS, Osservatorio di Astrofisica e Scienza dello Spazio di Bologna, via Gobetti 93/3, I-40129 Bologna, Italy}

\author[0000-0003-1096-2636]{Erik Zackrisson}
\affiliation{Observational Astrophysics, Department of Physics and Astronomy, Uppsala University, Box 516, SE-751 20 Uppsala, Sweden}

\author[0000-0002-0350-4488]{Adi Zitrin}
\affiliation{Physics Department, Ben-Gurion University of the Negev, P.O. Box 653, Be'er-Sheva 84105, Israel}



\begin{abstract}
We present results of \Cii$\,158\,\rm{\mu m}$ emission line observations, and report the spectroscopic redshift confirmation of a strongly lensed ($\mu\sim20$) star-forming galaxy, MACS0308-zD1 at $z=6.2078\pm0.0002$.
The \Cii\ emission line is detected with a signal-to-noise ratio $>6$ within the rest-frame UV bright clump of the lensed galaxy (zD1.1) and exhibits multiple velocity components; the narrow \Cii\ has a velocity full-width-half-maximum (FWHM) of $110\pm20\,\rm{km/s}$, while broader \Cii\ is seen with an FWHM of $230\pm20\,\rm{km/s}$.
The broader \Cii\ component is blueshifted ($-80\pm20\,\rm{km/s}$) with respect to the narrow \Cii\ component, and has a morphology which extends beyond the UV-bright clump.
We find that while the narrow \Cii\ emission is most likely associated with zD1.1, the broader component is possibly associated with outflowing gas.
Based on the non-detection of $\lambda_{\rm 158\,\mu m}$ dust continuum, we find that MACS0308-zD1's star-formation activity occurs in a dust-free environment with the stringent upper limit of infrared luminosity $\lesssim9\times10^{8}\,{\rm L_{\odot}}$. 
Targeting this strongly lensed faint galaxy for follow-up ALMA and JWST observations will be crucial to characterize the details of typical galaxy growth in the early Universe.
\end{abstract}



\section{Introduction} \label{sec:intro}
Over the past decades, optical/near-infrared surveys using ground- and space-based telescopes have built large samples of high-redshift galaxies at $z > 6$, revealing star-formation activity and stellar mass buildup in the first Gyr of the Universe \citep[e.g.,][]{Madau2014,Stefanon2021,Weaver2022}.
Recently, our understanding of early galaxy buildup has been significantly advanced by unprecedented sensitivity and high-resolution observations of the Atacama Large Millimeter/submillimeter Array (ALMA).
ALMA observations have built large samples of 
of high-redshift galaxies with strong far-infrared (FIR) emission lines (e.g., \Cii\,$158\,\rm{\mu m}$, \Oiii\,$88\,\rm{\mu m}$) and dust continuum \citep[e.g.,][]{Hashimoto2018,Bethermin2020,Fudamoto2021,Bouwens2022}, and ultimately revealing the interstellar medium (ISM) properties of these early-epoch galaxies.

 ALMA \Cii\,$158\,\rm{\mu m}$ emission line surveys have so far revealed ISM properties for $>100$ for galaxies at high redshift, including: galaxy-scale morpho-kinematics which suggest a large diversity of formation pathways among $z>4$ star-forming galaxies \citep[][]{lefevre2020,Jones2021}, $\sim$500--1000\,$\rm{km/s}$ gas outflows velocities \citep[][]{Maiolino2012,Ginolfi2020}, and $\sim$30\,$\rm{kpc}$ scale gas halos surrounding a dominant fraction of galaxies which imply a star-formation driven metal pollution of the circumgalactic medium \citep[][]{Fujimoto2020,Fudamoto2022}. 
A strong correlation between the neutral gas reservoir seen by \Cii\ and SFR of galaxies \citep[][]{Schaerer2020}. Evolution of dust-obscured star-formation activities suggests an evolution of dust attenuation properties at $z>4$ \citep[][]{Bouwens2020,Fudamoto2020}. It has been proposed to use \Cii\,$158\,\rm{\mu m}$ line as a tracer of star-forming neutral gas, which shows evolution of the neutral gas fraction of high-redshift galaxies \citep[][]{Zanella2018,Dessauges2020,Heintz2021}.

To date, these high-$z$ observations have been mostly limited to massive star-forming galaxies (e.g., $M_{\ast}>10^{10}\,\rm{M_{\odot}}$ and/or $\rm{SFR} > 10\,\rm{M_{\odot}\,yr^{-1}}$) and we only have small samples of less evolved, low-mass/low-SFR galaxies.
This is because detailed ISM observations of relatively low-mass galaxies are still one of the major challenges even for ALMA as they are intrinsically faint.
One method to push to fainter limits is to adopt strongly lensed galaxies to investigate the detailed properties of high-redshift, low-mass galaxies in a feasible observing time \citep[e.g.,][]{Coe2019,Treu2022,Vanzella2022}. Such strongly magnified high-redshift sources are, however, relatively rare and only a handful of ALMA observations have been performed so far \citep[][]{Knudsen2016,Bradac2017,Calura2021,Fujimoto2021}. The study of the ISM properties of low-mass, high-$z$ galaxies is thus still in its nascent stages.

Here, we report an observation of a strongly lensed  $\sim L^{\ast}$ star-forming galaxy: MACS0308-zD1 at $z\sim6.2$. We clearly detect a \Cii\,$158\,\rm{\mu m}$ emission line and measures the spectroscopic redshift of MACS0308-zD1 for the first time \citep{Acebron2018,Salmon2020,Welch2023}. Making use of high-resolution, high-sensitivity ALMA observations, we also study the velocity structure and morphology of the \Cii\ emission. These suggest multiple components of \Cii\ emission that are distinct in spatial and spectral directions.

This paper is organized as follows: in \S\ref{sec:obs} we describe our target galaxy and the ALMA observations used in this study. In \S\ref{sec:analysis}, we present our data analysis and measurements for the \Cii\,$\,158\,{\rm \mu m}$ emission line and $\lambda_{\rm rest}\sim158\,{\rm \mu m}$ dust continuum.  In \S\ref{sec:discussion}, we compare to previous studies and discuss our results. Finally, we conclude with the summary in \S\ref{sec:conclusion}.
Throughout this paper, we assume a cosmology with $(\Omega_m,\Omega_{\Lambda},h)=(0.3,0.7,0.7)$, and the Chabrier \citep{Chabrier2003} initial mass function (IMF), where applicable.
With these cosmological parameters, 1 arcsec corresponds to $5.6\,\rm{kpc}$ at $z=6.2078$.

\section{Observation and Data Reduction} \label{sec:obs}

\subsection{Target: MACS0308-zD1}
The target of the ALMA observations, MACS0308-zD1, was identified in the Reionization Lensing Cluster Survey (RELICS; \citealt{Coe2019}) in the background of the lensing cluster MACS0308+26. The galaxy was first identified as one of the triply imaged galaxies in the field at $z>6$ \citep{Acebron2018}, and our target is the brightest among the three images ($\rm{RA= 03:08:53.407}$, $\rm{Dec=+26:44:58.93}$, which originally given the ID of MACS0308-c4.3 in \citealt{Acebron2018}).
MACS0308-zD1 was found as the apparently UV-brightest $z>6$ galaxy having $m_{\rm AB}=23.0\,{\rm mag}$, and an estimated photometric redshift of $z_{\rm ph}\sim6.2 - 6.3$ \citep{Salmon2020}. \citet{Welch2023} further analyzed MACS0308-zD1 by forward modeling of the galaxy in the source plane and found an updated robust photometric redshift of $z_{\rm ph}=6.21$, and that MACS0308-zD1 consists of a bright clump: MACS0308-zD1.1, and a faint diffuse component. The bright clump was found to have a radius of $r=27\,\rm{pc}$ in its source plane, and to have an extremely high star-formation rate surface density of $\Sigma_{\rm SFR}\sim900\,{\rm M_{\odot}\,yr^{-1}\,kpc^{-2}}$. This $\Sigma_{\rm SFR}$ is the highest value ever found at $z\sim6$ star-forming galaxies \citep{Kennicutt2012}, suggesting that the clump could be an environment similar to the starburst galaxies.

\subsection{ALMA observations}
\begin{figure*}
    \centering
    \includegraphics[width=\textwidth]{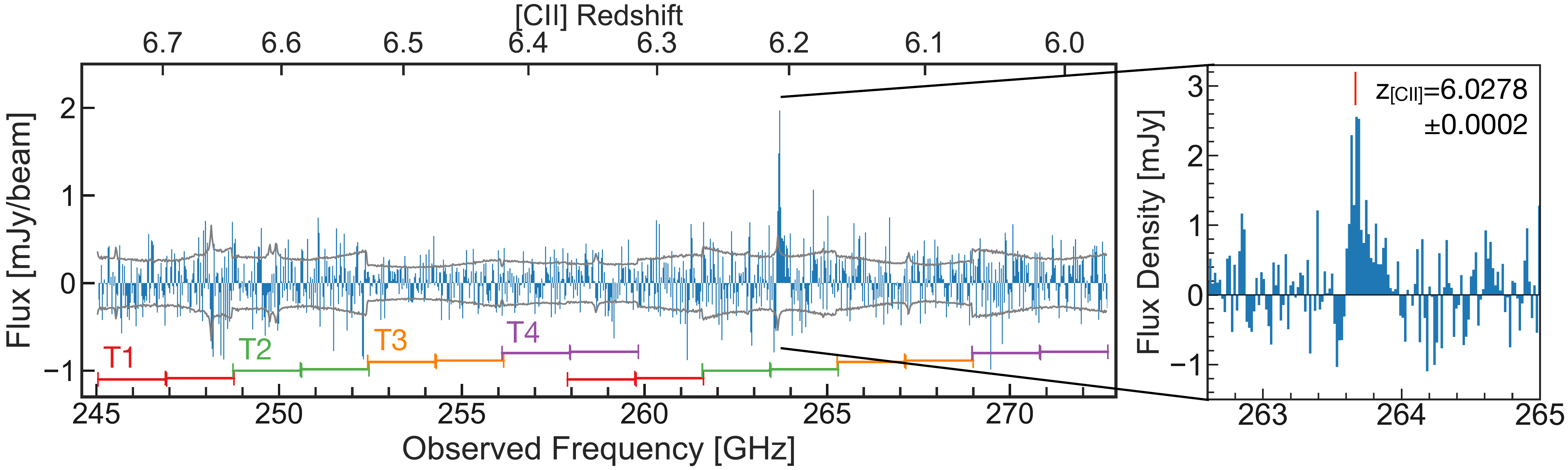}
    \caption{{\bf Left Panel:} The entire spectrum of the MACS0308-zD1 using a spectral bin width of $30\,\rm{km/s}$. The spectrum is extracted from a peak pixel of the line moment-0 map. Dust continuum was not subtracted as it is not detected. The gray line shows the pixel-by-pixel RMS of each channel. An emission line is significantly detected at the frequency of $263.700\pm0.004\,\rm{GHz}$ confirming a spectroscopic redshift of $z_{\rm [CII]}=6.2078\pm0.0002$. The bottom horizontal solid lines indicate the frequency coverage of each tuning (T1 to T4).
    {\bf Right Panel:} Zoom-in view of the spatially-integrated \Cii\ emission line of MACS0308-zD1 with a spectral bin width of $20\,\rm{km/s}$. The integrated spectrum was extracted using the $2\,\sigma$ area within its moment-0 map. The \Cii\ line has a skewed profile that suggests the existence of multiple velocity components in the \Cii\ emission (see \S\ref{sec:C2analysis}).
    }
    \label{fig:wholespec}
\end{figure*}

MACS0308-zD1 has been observed in the ALMA program \#2021.1.00143.S (PI: Fudamoto). The observation scanned to search the \Cii\,$\,158\,\rm{\mu m}$ emission line at a sky frequency range between $245\,\rm{GHz}$ and $273\,\rm{GHz}$ that covers entire photometric redshift probability of the target \citep{Acebron2018,Salmon2020, Welch2023}.
To efficiently scan the frequency range, the observations used four separate spectral tunings (T1 to T4; see Fig. \ref{fig:wholespec}). The correlator was set up in dual polarization and frequency domain mode, covering 1.875 GHz band width having a native spectral resolution of $\sim9\,\rm{km/s}$.

All tunings were observed during multiple observing runs using the ALMA cycle-8 C43-3 configuration. The on-source time was $\sim$0.5, $\sim$1.6 ,$\sim$1.0, and $\sim$1.0\,\rm{hrs}, respectively. The precipitable water vapor (PWV) during the observations was $0.545$, $1.450$, $0.544$, $0.636\,\rm{mm}$ for tunings T1, T2, T3, and T4, respectively. The relatively large variations of the on-source time were mainly due to variations in the weather conditions.

For the data calibration, we asked EA-ARC for the data calibration support\footnote{\url{https://www2.nao.ac.jp/~eaarc/DATARED/support_data_reduction_en.html}} to deliver the calibrated data that are produced using the QA2 pipeline code (i.e., \path{ScriptForPI.py}).
The observed data were analyzed using The Common Astronomy Software Application
(CASA) version 6.2.1 \citep{Bean2022}. 

\subsection{Imaging data cube and \Cii\ $158\,\rm{\mu m}$ line detection}

To search the \Cii\,$\,158\,\rm{\mu m}$ emission line, we created a data cube combining all four tunings using CASA task \path{tclean} using \texttt{NATURAL} weighting scheme,  a spectral binning of $30\,\rm{km/s}$, a pixel scale of $0.1^{\prime\prime}$, and an image size of $30^{\prime\prime}\times30^{\prime\prime}$ (i.e., a large enough sky area that contains the primary beam attenuation $<50\%$). With CASA task \texttt{tclean}, we performed the synthesized beam deconvolution using a fixed stopping threshold of three times the pixel-by-pixel root mean square (RMS) for all channels and the maximum  iteration of 500 times. The conservative deconvolution threshold does not affect the final data products as there are no bright sources in this velocity binning that could contaminate the resulting image products. The produced data cube has a synthesized beam full width at half maximum (FWHM) of $1.31^{\prime\prime}\times0.73^{\prime\prime}$, $0.94^{\prime\prime}\times0.68^{\prime\prime}$, $1.11^{\prime\prime}\times0.78^{\prime\prime}$, and $1.03^{\prime\prime}\times0.73^{\prime\prime}$ for tuning 1, 2, 3, and 4, respectively. The average pixel-by-pixel RMS was $0.27\,\rm{mJy/beam}$ for $30\,\rm{km/s}$ bin. The achieved sensitivity varies by a factor of up to $2.4$ from tuning to tuning. The variations in the sensitivity are due to the variations of the precipitable water vapor (PWV) under which each tuning was observed and also due to the weak atmospheric absorption features (see Fig. \ref{fig:wholespec}).

Using the location of the rest-frame UV emission as a prior of the spatial position of \Cii, we found the \Cii$\,158\,\rm{\mu m}$ emission line from MACS0308-zD1 at a frequency of $263.6995\pm0.004\,{\rm GHz}$ (Fig. \ref{fig:wholespec}).

\subsection{Imaging continuum map}
The $\lambda_{\rm rest}\sim158\,\rm{\mu m}$ continuum image of MACS0308-zD1 was created using CASA task \path{tclean} with the multi frequency synthesis mode. During the continuum imaging, we conservatively excluded the whole spectral window that contains the entire \Cii\,$\,158\,\rm{\mu m}$ emission line. Since we only excluded one spectral window out of the 16 total spectral windows from our four spectral setups, the exclusion does not affect the resulting sensitivity of the continuum map. 
The synthesized beam deconvolution was performed using \texttt{NATURAL} weighting scheme, using a stopping threshold of $27\,\rm{\mu Jy/beam}$ (i.e., $3\times$ the pixel-by-pixel RMS of the dirty image) and by setting the maximum number of iteration to 500 times. Again, the deconvolution thresholds do not affect the result, as there are no bright sources in the continuum image. As a result, we obtained a dust continuum image of MACS0308-zD1 that has a final pixel-by-pixel RMS of $8.9\,\rm{\mu Jy/beam}$ (Fig. \ref{fig:continuum}).

\section{Analysis and Measurements} \label{sec:analysis}

\subsection{Dust Continuum}

\begin{figure}
    \centering
    \includegraphics[width=0.55\columnwidth]{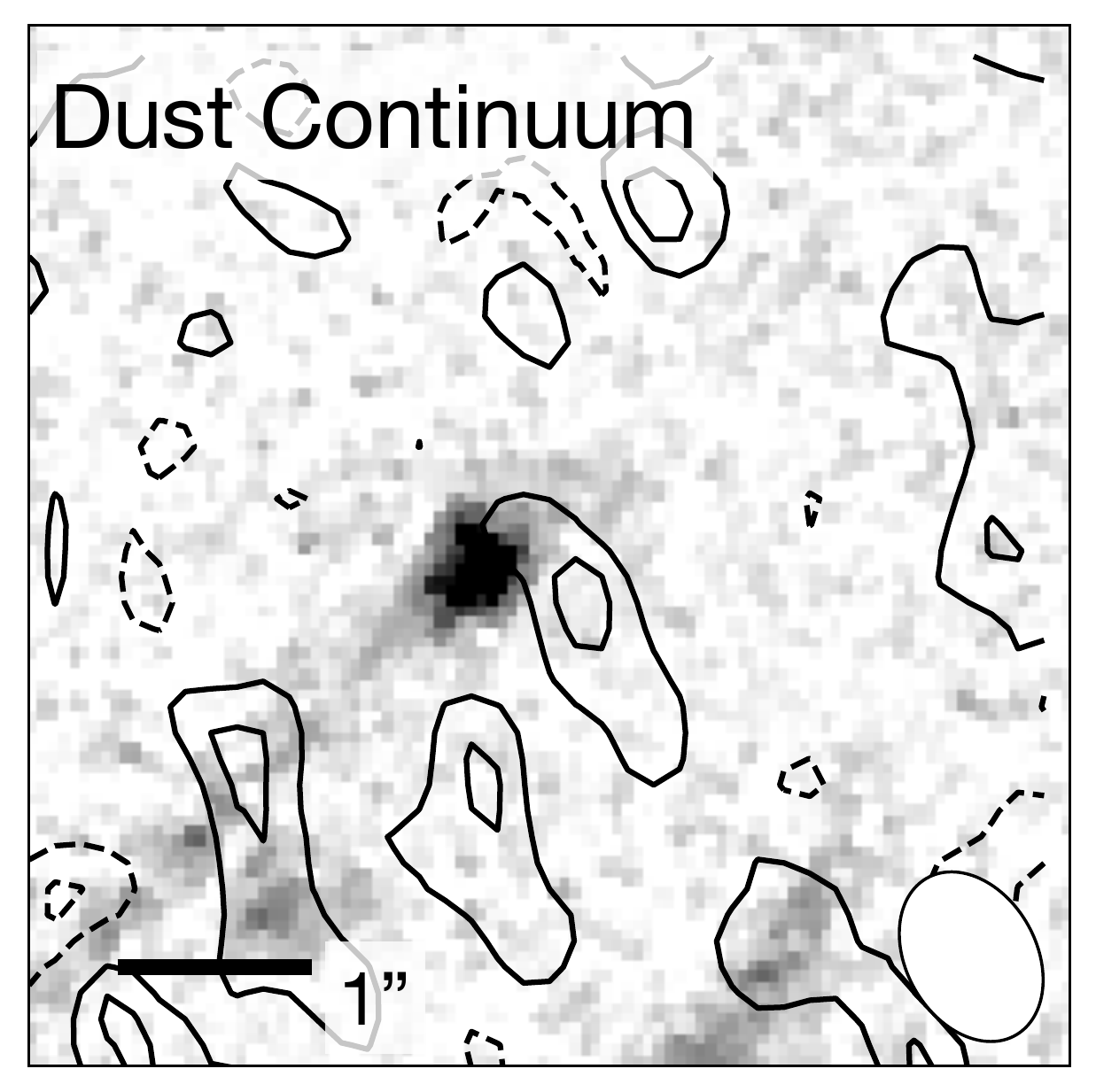}
    \caption{$6^{\prime\prime}\times6^{\prime\prime}$ cutout of HST F160W image (background) and dust continuum (Contours). Solid (dashed) contours show $1,2\,\sigma$ ($-2,-1\,\sigma$) signals of the dust continuum image. The astrometry of HST F160W is matched to the GAIA DR3 coordinates \citep{Gaia2022}. The white ellipse in the lower right corner shows the synthesized beam FWHM of the continuum image. As there is no significant signal spatially co-located with the rest-frame UV detections in the HST image, we conclude that the dust continuum of MACS0308-zD1 is non-detected in our observations, and we provided a flux $3\,\sigma$ upper limit of $<27\,\rm{\mu Jy}$.}
    \label{fig:continuum}
\end{figure}

In the continuum image, there is only a $\sim2\,\sigma$ positive signal close to but offset from the rest-frame UV position of the MACS0308-zD1 (Fig. \ref{fig:continuum}). Several previous studies found that typically a significance of $>3.5\,\sigma$ is required to avoid contamination of noise signals \citep[e.g.,][]{Bethermin2020,Inami2022}.
Therefore, we conclude that the dust continuum of MACS0308-zD1 is not detected in the observations.

As the dust continuum is not detected, we estimated the $3\,\sigma$ upper limit of the rest-frame $158\,\rm{\mu m}$ continuum of MACS0308-zD1 as $<27\,\rm{\mu Jy}$, adopting $3\times$ the pixel-by-pixel RMS.
Assuming a dust temperature of $\sim45\,\rm{K}$  and $\beta=1.8$ \citep{Bethermin2020, Sommovigo2022} for a modified black body emission, we estimated a $3\,\sigma$ upper limit of the observed infrared (IR) luminosity of $<1.9\times10^{10}\,\rm{L_{\odot}}$ in the image plane. This means $L_{\rm IR}<9.5\times10^{8}\,\rm{L_{\odot}}$ in the source plane assuming $\mu=20$.

\subsection{\Cii\ $158\,\rm{\mu m}$ emission line} \label{sec:C2analysis}

\begin{figure}
    \centering
    \includegraphics[width=1\columnwidth]{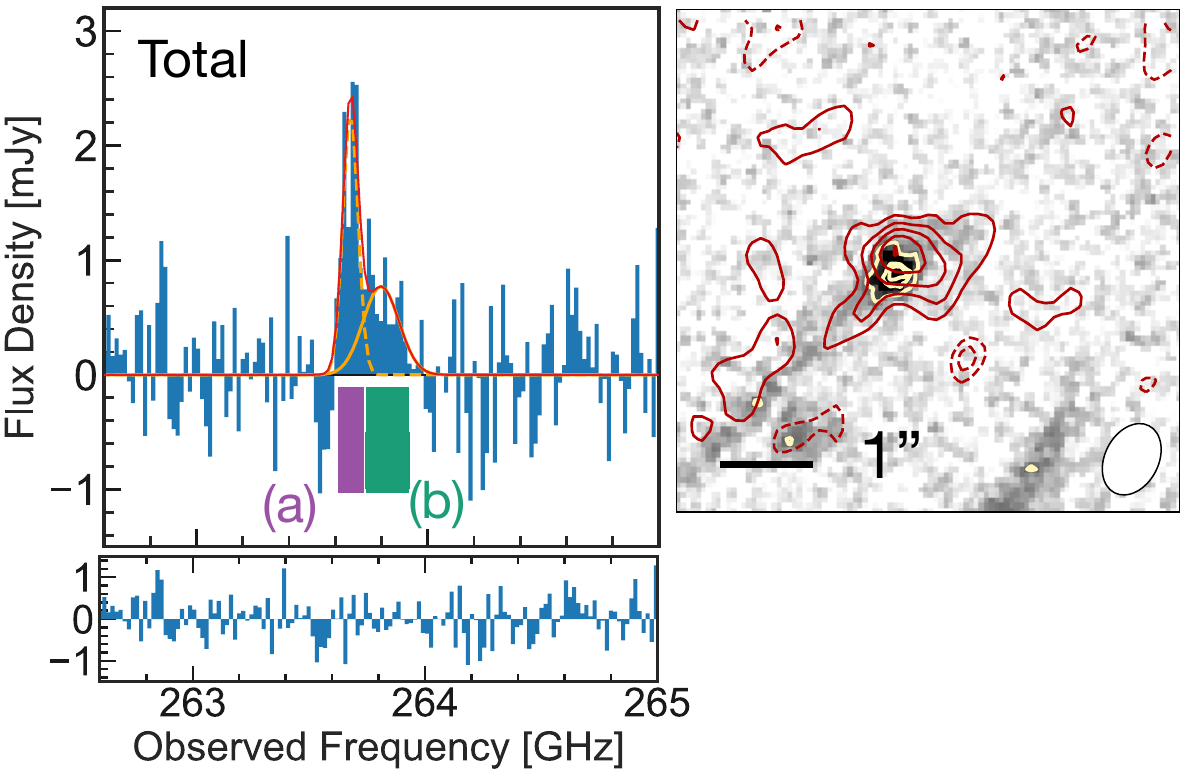}
    \caption{
    {\bf Left panel:} The \Cii$\,158\,\rm{\mu m}$ emission line integrated over entire components. 
    The red line shows the result of two-component 1-D Gaussian fitting to the spectra. Orange lines show individual Gaussians for narrow (dashed line) and broader (solid line) components.
    The residuals of the fits are indicated in the lower inset. The inset shows that there are no noticeable residuals by comparing to the rest of the frequency range. The frequency ranges indicated with the boxes (a) and (b) are used to make individual moment-0 images for the narrow and the broader components (\S\ref{sec:twocomp}, see also Fig. \ref{fig:mom0_separate}).
    {\bf Right panel:} $6^{\prime\prime}\times6^{\prime\prime}$ cutout of HST F160W image with \Cii\ moment-0 contours (red solid contours: 2 to 6 $\,\sigma$, red dashed contours: $-3,-2\,\sigma$). Yellow contours show $5,15,25\,\sigma$ of the F160W image. The moment-0 map is made by integrating $263.6$ to $263.99\,\rm{GHz}$. The white ellipse in the lower right corner shows the synthesized beam FWHM of the moment-0 image.
    Astrometry of the HST image is matched to the GAIA DR3 catalog \citep{Gaia2022}.
    }
    \label{fig:mom0_total}
\end{figure}

We extracted the \Cii\ emission line and created the moment-0 map in an iterative way. First, we created a moment-0 map integrating over an arbitrary frequency range that clearly include the \Cii\ emission line. Second, we extracted the \Cii\ spectrum using the moment-0 maps as a mask. Namely, we used pixels that had $>2\,\sigma$ signals in the moment-0 map, and extracted integrated spectra that were again used to create moment-0 map using channels having continuous positive signals. The procedures are repeated until the channels used to create moment-0 map converged. A few iterations were enough to converge, with the final results shown in Fig \ref{fig:mom0_total}.

In this way, we obtained the moment-0 image and the spectrum of the \Cii\ emission line integrated over the whole galaxy component.
Using the integrated moment-0 map, we measured the total \Cii\ flux density using CASA task \texttt{imfit}; 2-dimensional Gaussian fitting. Measured quantities are summarized in Tab. \ref{tab:measurements}.

\subsection{Multi-Component Analysis of the \Cii\ line} \label{sec:twocomp}

The \Cii\ emission line of MACS0308-zD1 deviates from a single Gaussian profile, and shows a blueshifted excess (Fig. \ref{fig:mom0_total}). We separated these components and estimated the properties of each of the \Cii\ lines.
As a first step, we performed a least square fitting using a two component Gaussian function to the integrated 1-D spectrum of the \Cii\ emission. This resulted in a reasonable fit with only small residuals that were consistent with random noise.
We found that the best fitting results arose from the combination of a narrow and a broader Gaussian (Left panel of Fig. \ref{fig:mom0_total}).

\begin{figure}
    \centering
    \includegraphics[width=1\columnwidth]{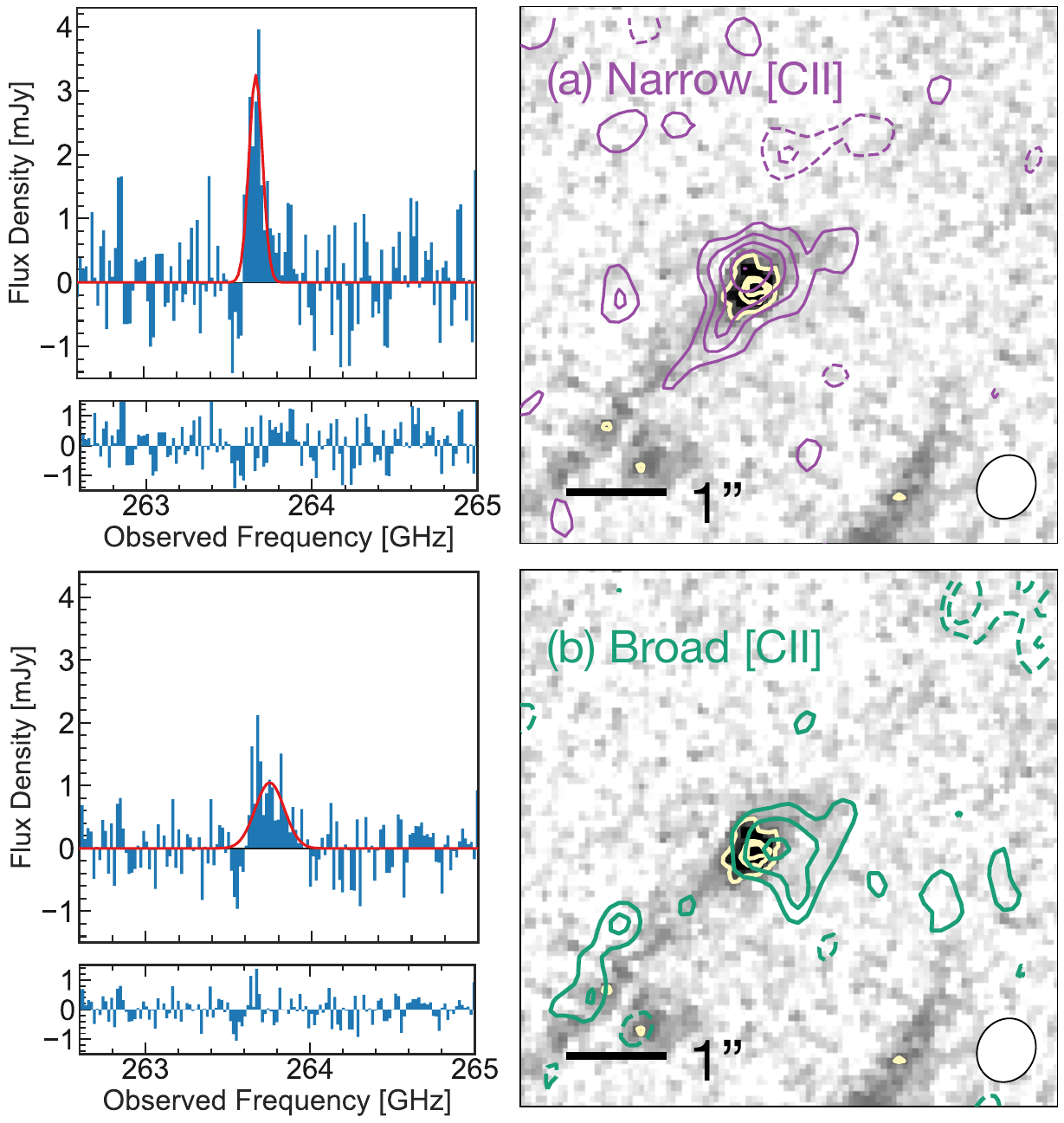}
    \caption{
    {\bf Left panels}: The extracted \Cii$\,158\,\rm{\mu m}$ spectrum using pixels with $>2\,\sigma$ significance in the moment-0 maps.
    The red lines show the results of single-component 1-D Gaussian fitting to the spectra. The residuals of the fits are indicated in the lower insets.
    {\bf Right panels} show $6^{\prime\prime}\times6^{\prime\prime}$ cutouts of HST F160W image (astrometry corrected using GAIA DR3; \citealt{Gaia2022}) with \Cii\ moment-0 contours (purple or green solid contours: 2 to 6 $\,\sigma$, purple or green dashed contours: $-3,-2\,\sigma$). Yellow contours show $5,15,25\,\sigma$ of the F160W image. The white ellipse in the lower right corner shows the synthesized beam FWHM of the moment-0 image.
    {\bf Top:} \Cii\ emission line integrated over the narrow component. The moment-0 map is made by integrating $263.62$ to $263.72\,\rm{GHz}$. {\bf Bottom:} Broader component of \Cii\ emission. The moment-0 map is made by integrating $263.73$ to $263.92\,\rm{GHz}$. While the narrow \Cii\ is spatially co-located with the HST F160W detection, the broader \Cii\ appears somewhat offset from the rest-UV emission.
    }
    \label{fig:mom0_separate}
\end{figure}

We then created moment-0 maps for both the narrow and the broader \Cii\ emission by integrating frequency ranges of $263.62$ to $263.72\,\rm{GHz}$ and $263.73$ to $263.92\,\rm{GHz}$, respectively. These frequency ranges were selected to separately image moment-0 maps of the two different components, and at the same time, to avoid contamination of different components.

Based on the results of the two component Gaussian fitting, we estimated the fractions of contamination for each maps.
Using the above frequency range to integrate and the results of the two component Gaussian fitting, we estimated that $\sim7\,\%$ of the narrow \Cii\ component contaminates the broad image, while $\sim8\%$ of the broader \Cii\ component contaminates the narrow \Cii\ moment-0 maps. Thus, the contamination from other components has only a minor impact on the moment-0 maps and measured fluxes for each component.

\subsubsection{The Narrow \Cii\ Emission Component}

The moment-0 image of the narrow \Cii\ component shows a $6\,\sigma$ detection. The emission is co-spatial with the rest-frame UV emission of the bright clump (hereafter MACS0308-zD1.1) seen by the HST F160W filter (top panel of Fig. \ref{fig:mom0_separate}). The significant detection from the co-spatial location implies that the narrow component is associated with the UV-bright star-formation activity of the MACS0308-zD1.1.
We extracted the narrow component of the \Cii\ emission using the moment-0 map (top left panel of Fig. \ref{fig:mom0_separate}), which has a velocity FWHM of $110\pm19\,\rm{km/s}$.
This line width leis at the lower bound of the FWHM of \Cii\ emission lines thus far observed from high redshift galaxies \citep[e.g.,][]{Bethermin2020}.

\subsubsection{The Broader \Cii\ Emission Component} \label{sec:broadc2}

The moment-0 image of the broader \Cii\ shows a $\sim$4.2\,$\sigma$ detection. The emission appears somewhat offset from rest-frame UV emission of the MACS0308-zD1.1 (bottom panel of Fig. \ref{fig:mom0_separate}).
The extracted emission of the broader \Cii\ component has a velocity FWHM of $233\pm19\,\rm{km/s}$, roughtly two times broader than the narrow component.  The broader component appears blueshifted by $\Delta v = -80\pm20\,\rm{km/s}$ with respect to the narrow component (Tab. \ref{tab:measurements}).

\renewcommand{\arraystretch}{1.2}
\begin{table*}
\begin{threeparttable}
\caption{Summary of the ALMA measurements}
\label{tab:measurements}
    \centering
    \begin{tabular}{lllllll}
    \hline\hline
     \Cii\ component  & Redshift & \Cii\ FWHM & $f_{\rm [CII]} \times \mu^{\dagger}$ & $L_{\rm [CII]}$ & $M_{\rm H_2}$\\
     & & $\rm{km/s}$ & ${\rm Jy\,km/s}$ & ${\rm L_{\odot}}$ & ${\rm M_{\odot}}$\\
     \hline
     Total    & $6.2078\pm0.0002$ & -- & $0.62\pm0.18$ & $2.96\pm0.79\times 10^{7}$ & $9.05\pm0.81\times 10^{8}$\\
     Narrow & $6.2078\pm0.0002$ & $110\pm20$ & $0.37\pm0.10$ & $1.67\pm0.45\times10^{7}$ & $5.05\pm0.45\times 10^{8}$\\
     Broad & $6.2058\pm0.0005$ & $230\pm20$ & $0.30\pm0.09$ & $1.35\pm0.45\times10^{7}$ & $4.05\pm0.40\times 10^{8}$\\
     \hline
    \end{tabular}
    \begin{tablenotes}
	\item  We used the single component 2D Gaussian fits to measure integrated fluxes.\vspace{-0.4cm}\\
        \item   $\dagger$ The measured values (except for the \Cii\ flux densities) are corrected for gravitational lensing magnification of $\mu=22$.
\end{tablenotes}
\end{threeparttable}
\end{table*}

\section{Discussion} \label{sec:discussion}

\subsection{Dust-obscured Star-Formation Activity} \label{sec:continuum}

We derive a $3\,\sigma$ upper limit to the observed IR luminosity of $<1.9\times10^{10}\,\rm{L_{\odot}}$ assuming a dust temperature of $45\,\rm{K}$.
Assuming that the dust emission is co-spatial with MACS0308-zD1, the $3\,\sigma$ upper limit IR luminosity in the source plane is $\lesssim9\times10^{8}\,\rm{L_{\odot}}$, applying a magnification factor of $\mu\sim20$ \citep{Welch2023}. The stringent upper limit on the IR luminosity implies an extremely low dust-obscured star-formation rate of ${\rm SFR_{IR}}<0.1\,\rm{M_{\odot}\,yr^{-1}}$ using the conversion of ${\rm SFR_{IR}}=1.2\times10^{-10}\,L_{\rm IR}\,{\rm M_{\odot}\,yr^{-1}\,L_{\odot}^{-1}}$ \citep{Madau2014,Inami2022}. Given the extreme brightness of MACS0308-zD1 in the rest-frame UV ($SFR_{\rm UV}=3\pm1\,\rm{M_{\odot}\,yr^{-1}}$ for zD1.1 and $7\pm2\,\rm{M_{\odot}\,yr^{-1}}$ for diffuse component; \citealt{Welch2023}), we derive a very low dust-obscured fraction of the star formation activity ($f_{\rm obs} = \rm{SFR_{IR}/\rm{SFR_{total}}}$) of only $\lesssim 1$ to $ 3\,\%$, meaning that the star formation activity of MACS0308-zD1 is almost completely dust-free.

The major uncertainty in estimating the IR luminosity from a single ALMA observation is the unknown dust temperature; increasing the assumed dust temperature can produce larger IR luminosities \citep[e.g.,][]{Faisst2020,Bakx2021,Fudamoto2022Td,Algera2023}. We examined this effect and its impact on the dust-obscured SFR. For a high dust temperature case of $60\,\rm{K}$, the de-magnified $3\,\sigma$ upper limit IR luminosity is $5.5\times10^{9}\,\rm{L_{\odot}}$ which yields a $3\,\sigma$ upper limit SFR of ${\rm SFR_{IR}}<0.3\,\rm{M_{\odot}\,yr^{-1}}$. Even for an extreme dust temperature of $80\,\rm{K}$, we find $8.0\times10^{9}\,\rm{L_{\odot}}$ and $0.8\,\rm{M_{\odot}\,yr^{-1}}$ for de-magnified $3\,\sigma$ upper limit of IR luminosity and dust-obscured SFR, respectively. Although such extremely high dust temperatures are not yet confirmed in this redshift range \citep[e.g.,][]{Algera2023}, even with these extreme assumptions, we still derived low obscured fractions of star-formation activity of $\lesssim10\,\%$ to $26\,\%$ for zD1.1 and the diffuse component, respectively. Thus, MACS0308-zD1 has relatively little dust-obscured activity, even after accounting for rather extreme uncertainties in the dust temperature.

This low $f_{\rm obs}$ contrasts strongly with the relatively high obscured fractions ($\sim50\,\%$) seen for massive $M_{\ast}>10^{9.5}\,\rm{M_{\odot}}$ galaxies at $z\gtrsim5$ \citep{Fudamoto2020,Schouws2022, Algera2023}. Nonetheless, it is consistent with the very blue UV color of MACS0308-zD1.1 \citep{Welch2023}. The nearly dust-free environment of the star-forming region is indicative of the metal-poor and young environment of the star-formation activity. In such an environment, strong \Oiii\,$88\,\rm{\mu m}$ emission line is expected \citep[i.e., $L_{\rm [OIII]}/L_{\rm [CII]} \gtrsim 3$; ][]{Inoue2014,Hashimoto2019,Sugahara2022}. 
Further follow-up observations with ALMA will be important to characterize conditions of the interstellar medium of the star-forming region of the MACS0308-zD1.

\subsection{\Cii$\,158\,\rm{\mu m}$ Emission Line}

\begin{figure}
    \centering
    \includegraphics[width=0.9\columnwidth]{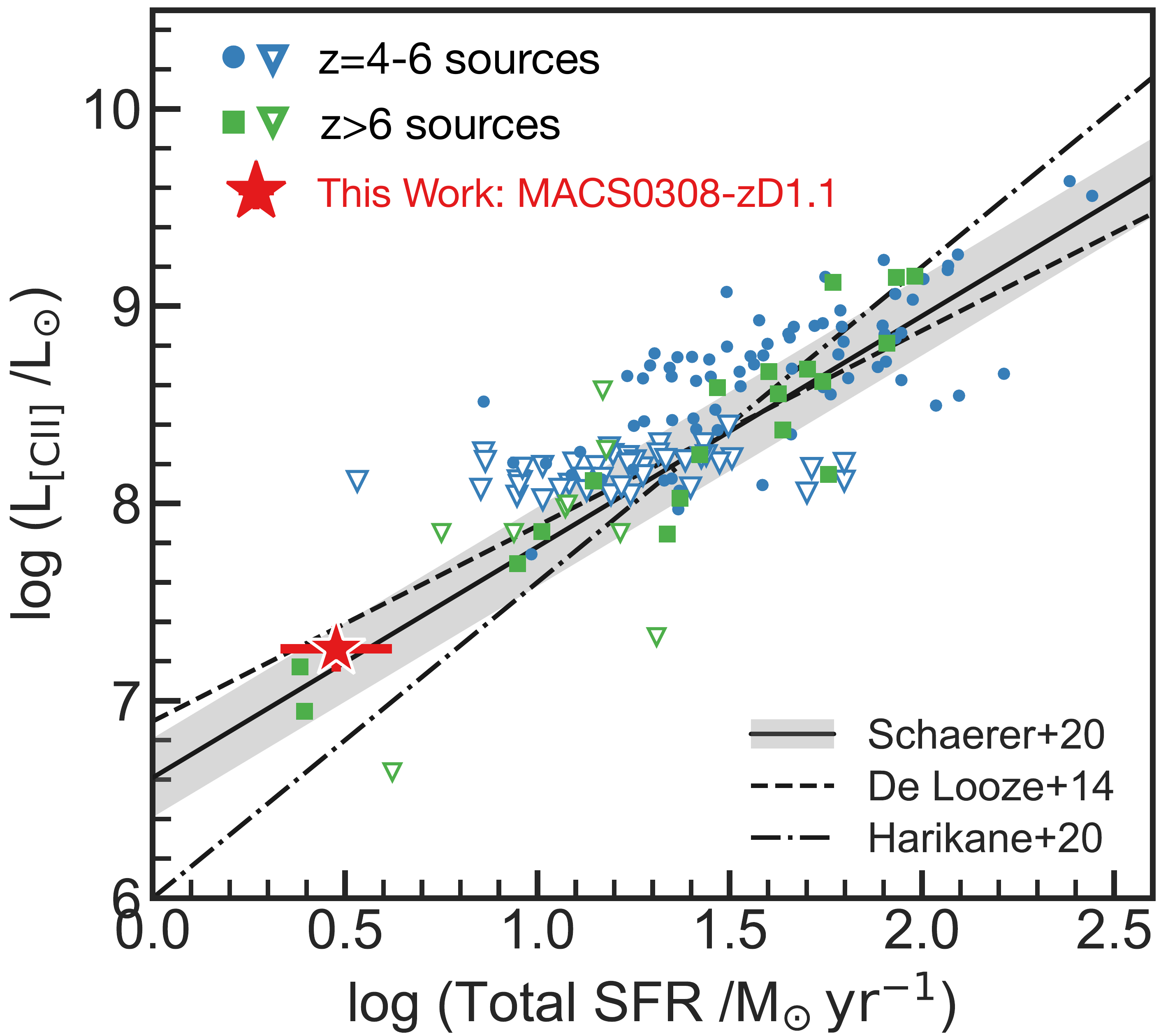}
    \caption{Observed \Cii\ luminosities versus star formation rates of galaxies at $4<z<6$ (blue points and triangles; \citealt{Schaerer2020} and references therein) and $z>6$ (green squares and triangles; \citealt{Matthee2019,Harikane2020} and references therein). Downward triangles show $3\,\sigma$ upper limits of the \Cii\ non-detections. SFRs are estimated using SED fitting or dust attenuation corrections \citep[i.e., the IRX-$\beta$ relation;][]{Fudamoto2020}. Lines show previously derived relations for low- and high-redshift galaxies (solid: \citealt{Schaerer2020}, dashed; \citealt{delooze2014}, and dot-dashed; \citealt{Harikane2020}). The \Cii\ luminosity of MACS0308-zD1 remains consistent with the locally calibrated relation and the relation observed for $z\sim5$ galaxies \citep{delooze2014,Schaerer2020}, but lies well above from the relation with ``[CII]-deficit'' galaxies \citep{Harikane2020}.} 
    \label{fig:sfrc2}
\end{figure}

\subsubsection{SFR-\Cii\ correlation}
It is highly debated whether the correlation between star formation rates (SFRs) and \lcii\ evolves as a function of redshift. Previous studies have shown that galaxies with low SFRs or metallicities show extremely faint \Cii\ emission, denoted ``\Cii-deficit'' \citep[e.g.,][]{Matthee2019,Harikane2020}, that are also predicted by theoretical studies \citep[e.g.,][]{Vallini2015, Liang2023}. However, \citet{Schaerer2020} demonstrated that, based on large samples of galaxies obtained in the ALMA large program ALPINE \citep{lefevre2020, Bethermin2020,Faisst2020}, there is no significant change of the observed \lcii-SFR relation as a function of redshift, and pointed out that the previously obtained ``\Cii-deficit'' is largely affected by the uncertain estimations of the dust-obscured SFRs of high-redshift galaxies.

With the measured \Cii\ luminosity of MACS0308-zD1.1 and estimated SFR ($3\pm1\,\rm{M_{\odot}\,yr^{-1}}$), MACS0308-zD1.1 appears consistent with the SFR-$L_{\rm [CII]}$ relation derived in the local galaxies \citep[e.g.,][]{delooze2014} and for high-redshift galaxies where no-evolution of the relation is found \citep{Schaerer2020}, but deviates from the ``\Cii-deficit'' relation suggested in some of the previous studies \citep{Harikane2020, Liang2023}. 
Here, we assume that the $\rm{SFR_{IR}}$ is negligible, given the faint IR luminosity upper limits and low SFR$_{\rm IR}$s found in \S\ref{sec:continuum}. 
Thus, our observation supports a non-evolving picture for the SFR-$L_{\rm [CII]}$ correlation in the star-forming clump-scale structure inside distant galaxies (Fig. \ref{fig:sfrc2}). This result suggests that the SFR-$L_{\rm [CII]}$ holds even for low SFR and potentially metal-poor galaxies at high redshift.

\subsubsection{Molecular Gas Mass} 
Previous studies showed that the \Cii$\,158\,\rm{\mu m}$ emission line is an excellent indicator of star-forming molecular gas.
In particular, \citet{Zanella2018} showed a tight correlation between molecular gas masses (${M_{\rm H2}}$) and \lcii\ in $z{\sim}2$ star-forming galaxies. Also, \cite{Dessauges2020} examined whether the ${M_{\rm H2}}$-\lcii\ relation is applicable to $4.5<z<5.5$ star-forming galaxies, and showed that the derived molecular gas masses are consistent with other methods.
Using the method presented in \citet{Zanella2018}, we estimated separately the molecular gas masses for the integrated, narrow \Cii\ and broad \Cii\ components of the MACS0308-zD1 (Tab. \ref{tab:measurements}).

We also estimated the surface density of molecular gas mass $\Sigma_{\rm H2}$. However, with the $\rm{FWHM}\sim1^{\prime\prime}$ resolution of the observation, both components of the \Cii\ emission are either unresolved or only marginally resolved. 
The most extreme case is found by assuming that the narrow \Cii\ is associated with the UV-bright component MACS0308-zD1.1, and assuming the size of the star-forming region is $r=27\,\rm{pc}$ \citep{Welch2023}. In this case, we found a remarkably dense surface gas mass density of ${\rm log}\,[\Sigma_{\rm H2}\,{\rm/(M_{\odot}\,pc^{2})}]=5.34\pm0.09$, which is $>6000$ times larger than that of giant molecular clouds in today's ISM \citep[e.g.,][]{Leroy2008}, and $\times100$ higher than the maximum value found in the recent simulation of galaxies at $z\sim8$ \citep{Garcia2022}.

Combined with the extremely high surface star formation rate density of $\Sigma_{\rm SFR}\sim900\,{\rm M_{\odot}\,yr^{-1}\,kpc^{-2}}$ \citep{Welch2023}, MACS0308-zD1.1 is located on the extreme edge of the Kennicutt-Schmidt relation \citep[e.g.,][]{Kennicutt2012}. This might suggest that the UV-bright clump has a star formation environment similar to or more extreme than starburst galaxies.

\subsection{\Cii\ outflow or another structure?}

\begin{figure}
    \centering
    \includegraphics[width=1.\columnwidth]{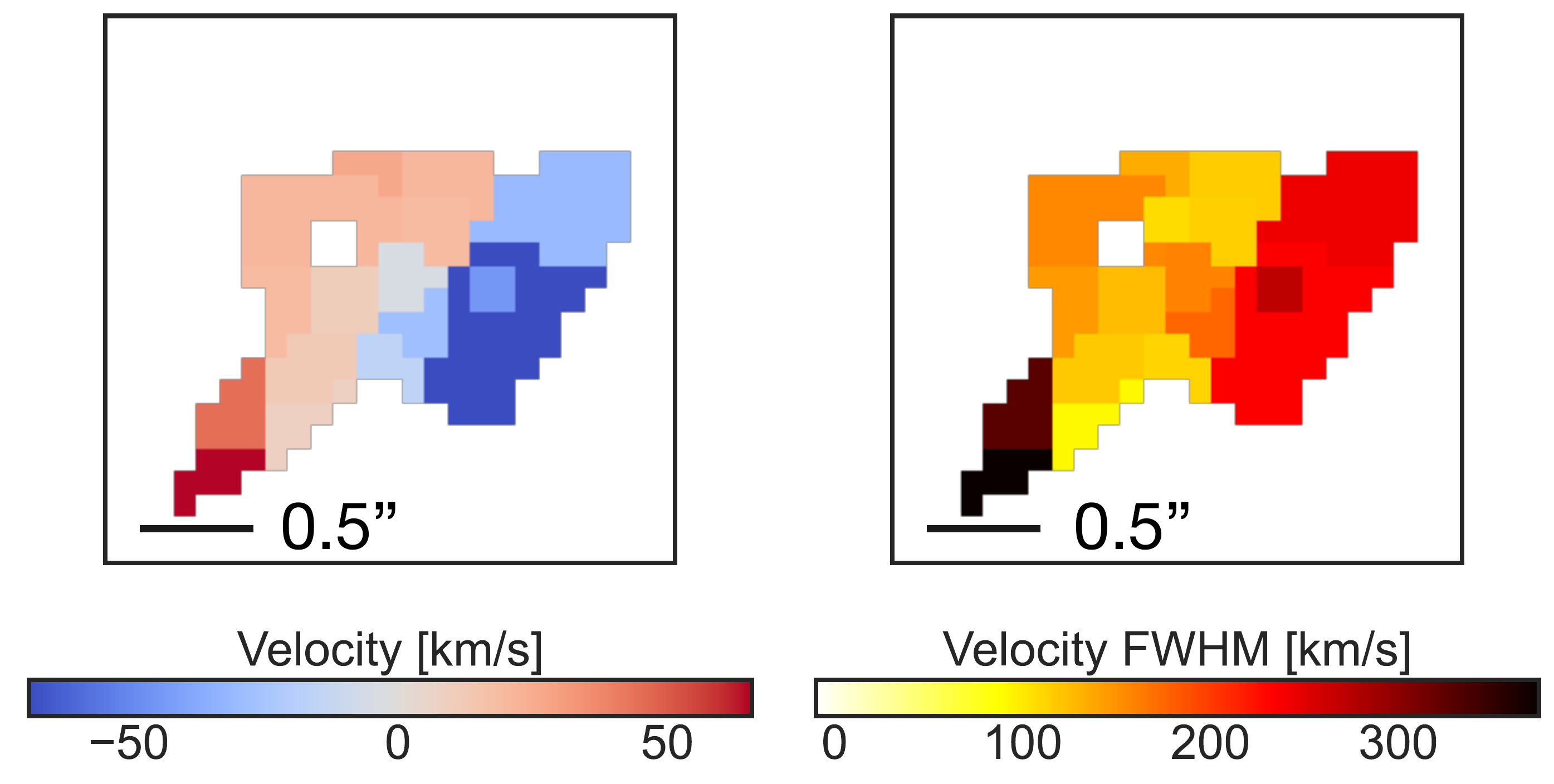}
    \caption{
    {\bf Left panel:} Velocity map of the \Cii\ emission line. The velocity zero point is defined by the redshift of $z=6.2078$.
    {\bf Right panel:} Velocity FWHM map of the \Cii\ emission line. The velocity and FWHM maps are made by 1D Gaussian fitting to the spectra extracted from the moment-0 map image using pixels above $>2\sigma$ signals. The measured velocity difference and FWHM matches well with the spatially-integrated measurement (Tab.\ref{tab:measurements}). Both the velocity and the FWHM map shows a gradient toward the same direction, suggesting that the simple rotation kinematics does not explain multiple components of the \Cii.}
    \label{fig:kinematics}
\end{figure}

We found multiple velocity components in the \Cii\ emission line: a narrow FWHM $\sim110\,\rm{km/s}$ \Cii\ that is spatially coincident with the UV-bright star-forming region, and a broader FWHM $\sim230\,\rm{km/s}$ \Cii\ that is spatially extended and potentially offset (Tab. \ref{tab:measurements}). 
The offset both in centroid velocity and dispersion in the same region is difficult to explain with simple rotation (Fig. \ref{fig:kinematics}).

As the narrow emission is spatially coincident with the UV emission of the galaxy, we associate it with the star-formation activity of the UV-bright clump, MACS0308-zD1.1. Assuming a gas mass of $M_{\rm H2}\,{\rm/M_{\odot}}=5.1\times10^{8}$ and a radius of $r=27\,{\rm pc}$, the line-of-sight velocity dispersion of a gas cloud will be $\sqrt{{\rm G} M/3r} = 167\,\rm{km/s}$, assuming virial equilibrium, where $\rm{G}$ is the gravitational constant. This value is consistent with the measured line width of the narrow component, suggesting that the ISM is supported by internal turbulent motion of the gas cloud.

On the other hand, the origin of the broad \Cii\ emission is less clear.
The broad line width and the velocity offset suggest that the broad \Cii\ emission line could arise from a physically distinct region from the narrow \Cii\ line. Additionally, the spatial extent of the broad \Cii\ component shows that the broad \Cii\ is spatially offset from the UV-bright star-forming region and the \Cii\ emission associated with the UV-bright clump. Although the estimated virial velocity of the cloud is uncertain without a confident size measurement, the fact that the linewidth of the broad component appears super-virial suggests that this component of the gas cloud might not be bound.
Combined with the extremely high density of star formation activity ($\Sigma_{\rm SFR}\sim900\,{\rm M_{\odot}\,yr^{-1}\,kpc^{-2}}$), circumstantial evidence suggests that the broad \Cii\ might arise from gas outflowing from the intense star-forming region.

An alternative interpretation of the broad \Cii\ component would be gas clumps generated as a result of the fragmentation of cold gas streams (i.e., inflow). Previous studies predicted such fragmentation of the cold gas streams in the inner halo ($\sim0.2\,R_{\rm v}$), where $R_{\rm v}$ is the virial radius \citep[e.g.,][]{Pallottini2017,Mandelker2018}. After the fragmentation of cold gas streams, the masses of each clump are predicted to be order of $\sim10^{5}$ to $10^{7}\,{\rm M_{\odot}}$. The estimated gas mass from our observation ($M_{\rm H2}\sim4\times10^{8}\,{\rm M_{\odot}}$) suggest that several clumps exist along the line of sight or the case that we are observing "down the barrel" of the stream, which increase observed gas mass and velocity width of the line.

More quantitative examination requires several follow-up observations, such as higher resolution and higher sensitivity ALMA \Cii\ follow-up and JWST's rest-frame optical emission to reveal the stellar distribution down to much deeper limits. Also, to examine the extremely dense environment of the star-forming ISM, ALMA observations using higher resolution or observations of a neutral dense gas indicator such as \Oi\ will be needed.

\section{Conclusion} \label{sec:conclusion}

In this paper, we presented results of \Cii$\,158\,\rm{\mu m}$ emission line scan observations of a strongly lensed ($\mu=22$) star-forming galaxy at $z\sim6.2$. We detected the \Cii\ emission line within the expected frequency range. Additionally, no dust continuum was detected. Based on the observations, we found the following:

\vspace{0.15cm}
\noindent$\bullet$ A spectroscopic redshift of $z=6.2078\pm0.0002$ was determined, and the result is consistent with previous photometric redshift estimates \citep{Salmon2020,Weaver2022}.

\vspace{0.15cm}
\noindent$\bullet$ The non-detection of the dust continuum emission suggests that MACS0308-zD1 is an almost dust-free environment.
In particular, the UV-bright clump, MACS0308-zD1.1 has little dust-obscured SFR of $<0.1\,\rm{M_{\odot}\,yr^{-1}}$, and a low dust-obscured fraction of $<3\%$.
The low dust-obscured activity is in stark contrast compared with the observations of other relatively massive $M_{\ast}>10^{9.5}\,\rm{M_{\odot}}$ star-forming galaxies \citep{Fudamoto2020,Schouws2022}, and suggest that the stellar populations in this galaxy are young and metal-poor.

\vspace{0.15cm}
\noindent$\bullet$ From the relatively bright detection of the \Cii\ emission line, we found that MACS0308-zD1.1 is consistent with the non-evolving SFR-\lcii\ correlation \citep{delooze2014,Schaerer2020}. Also, the bright \Cii\ emission suggests that the star-forming region of the galaxy may have an extremely dense molecular gas environment of ${\rm log}\,[\Sigma_{\rm H2}\,{\rm/(M_{\odot}\,pc^{2})}]=5.34\pm0.09$. With the extremely high density of star formation activity of $\Sigma_{\rm SFR}\sim900\,{\rm M_{\odot}\,yr^{-1}\,kpc^{-2}}$, the UV-bright star-forming region of MACS0308-zD1.1 is located in the most extreme edge of the Kennicutt-Schmidt relation \citep{Kennicutt2012}.

\vspace{0.15cm}
\noindent$\bullet$
We found that the \Cii\ emission line can be decomposed into two components; the narrow line ($\rm{FWHM}=110\pm19\,\rm{km/s}$) and the broad line ($\rm{FWHM}=233\pm19\,\rm{km/s}$). 
The broad component is blueshifted with respect to the narrow component with a velocity offset of $\Delta v = -80\pm20\,\rm{km/s}$.
Although higher spatial resolution (e.g., matched to the HST/JWST resolutions of $\sim0.16^{\prime\prime}$) and higher sensitivity observations are required to conclude, the narrow component is most likely associated with a gravitationally bound gas cloud with intense star forming activity, as seen by the UV-bright clump, and the broad component may belong to kinematically and spatially distinct activity in the galaxy (e.g., possibly outflowing gas).

\vspace{0.15cm}
This spectroscopic redshift confirmation of a strongly lensed $L^{\ast}$ star-forming galaxy MACS0308-zD1 at $z=6.2078$ demonstrates viability of the future follow-up ALMA observations.
Furthermore, JWST's NIRSpec IFU observation will be critical to understand the morphologically and kinematically complex structure of this galaxy. With the large magnification of $\mu\sim20$, MACS0308-zD1 will serve as a unique test case to study low-SFR and low-mas galaxy growths in great detail in the future observations.

\begin{acknowledgements}
This paper makes use of the following ALMA data: ADS/JAO.ALMA \#2021.1.00143.S ALMA is a partnership of ESO (representing its member states), NSF (USA) and NINS (Japan), together with NRC (Canada), MOST and ASIAA (Taiwan), and KASI (Republic of Korea), in cooperation with the Republic of Chile. The Joint ALMA Observatory is operated by ESO, AUI/NRAO and NAOJ. In addition, publications from NA authors must include the standard NRAO acknowledgement: The National Radio Astronomy Observatory is a facility of the National Science Foundation operated under cooperative agreement by Associated Universities, Inc. YF, YS, and AKI acknowledge support from NAOJ ALMA Scientific Research Grant number 2020-16B. PAO acknowledges support from the Swiss National Science Foundation through project grant 200020\_207349. The Cosmic Dawn Center (DAWN) is funded by the Danish National Research
Foundation under grant No. 140. EZ acknowledges funding from the Swedish National Space Agency and grant 2022-03804 from the Swedish Research Council (Vetenskapsr{\aa}det). MB acknowledges support from the Slovenian national research agency ARRS through grant N1-0238. A.Z. acknowledges support by grant No. 2020750 from the United States-Israel Binational Science Foundation (BSF) and grant No. 2109066 from the United States National Science Foundation (NSF), and by the Ministry of Science \& Technology, Israel. FEB acknowledges support from ANID-Chile BASAL CATA FB210003, FONDECYT Regular 1200495 and 1190818, and Millennium Science Initiative Program -- ICN12\_009.  

\end{acknowledgements}

%

\vspace{5mm}
\facilities{HST(STIS), ALMA}


\software{astropy \citep{astropy2022}
          }

\bibliography{base}{}
\bibliographystyle{aasjournal}



\end{document}